\documentclass[11pt]{article}

\usepackage{amsmath,amssymb}
\usepackage{graphicx}
\usepackage{epstopdf}

\newcommand{\be}{\begin{equation}}
\newcommand{\ee}{\end{equation}}
\newcommand{\bea}{\begin{eqnarray}}
\newcommand{\eea}{\end{eqnarray}}
\newcommand{\ba}{\begin{array}}
\newcommand{\ea}{\end{array}}

\numberwithin{equation}{section}

\newcommand{\vev}[1]{\left< #1 \right> }
\newcommand{\nn}{\nonumber}

\begin{document}

\allowdisplaybreaks

\title{ Size scaling of self gravitating polymers and strings \bigskip}

\author{
        Shoichi Kawamoto$^*$ 
        and
	Toshihiro Matsuo$^\dagger$\bigskip
	\\
	$^*$\footnotesize\it 
Department of Physics, Chung-Yuan Christian University,
Chung-Li 320, Taiwan, R.O.C.\\
	\footnotesize\tt kawamoto@cycu.edu.tw, kawamoto@yukawa.kyoto-u.ac.jp
	\smallskip\\
	$^\dagger$\footnotesize\it National Institute of Technology, Anan College, Tokushima 774-0017, Japan\\
	\footnotesize\tt matsuo@anan-nct.ac.jp, tmatsuo@yukawa.kyoto-u.ac.jp
	}

\date{\today}

\maketitle

\bigskip

\begin{abstract}
\noindent\normalsize

We study a statistical ensemble of a single polymer with self gravitational interaction.
This is a model of a gravitating string --- the precursor of a black hole.
We analyze averaged sizes by mean field approximations with an effective Hamiltonian \'a la Edwards with Newtonian potential as well as a contact repulsive interaction.
We find that there exists a certain scaling region where the attractive and the repulsive forces balance out.
The repulsive interaction pushes the critical gravitational coupling 
to a larger value, at which the size of a polymer becomes comparable to its Schwarzschild radius, and as a result the size of the corresponding black hole increases considerably.
We show phase diagrams in various dimensions that clarify how the size changes as
the strengths of repulsive and gravitational forces vary.

\end{abstract}

\vfill

\setcounter{footnote}{0}
\section{Introduction}

It has been long known that a typical configuration of a highly excited
free fundamental string has a size $R \simeq \ell_s \mathcal{N}^{1\over 4}$, where $\mathcal{N} (\gg 1)$ is the excited level and $\ell_s$ is the string length scale.
Since the mass of the level $\mathcal{N}$ states is proportional to
$\sqrt{\mathcal{N}}$, which may be identified with a typical length
of the string, the profile can be understood as
that of a free random walk \cite{Mitchell-Turok}.\footnote{%
Ma\~nes has studied a Rutherford-type scattering
for a free long string and provided a further evidence that the
profile
is indeed that of a free random walk \cite{Manes}.}
The size of the configuration will decrease if the interaction between each part of a long string is turned on.
At a sufficiently strong coupling, the size would be the string scale, 
and at that point Susskind conjectured that the
collapsed configuration of a string may be identified with a small black hole of the same energy
\cite{Susskind:1993ws, Horowitz:1996nw}.
A natural question to be addressed is how the size of a long string changes as the string coupling constant varies.
Horowitz and Polchinski \cite{Horowitz:1997jc} studied this problem by employing so-called thermal scalar theory, which provides the statistical nature of a long string near the Hagedorn temperature
such as a spatial profile of the string.\footnote{%
Properties of the thermal scalar are further investigated in \cite{Barbon:2004dd},
and recently more intensive studies have been carried out by 
Mertens et al. \cite{Mertens}.
There is another trial to evaluate it through the change of the density of states due to self interaction \cite{Damour:1999aw}.}
Through a scaling argument, they estimated a typical size of a bound
state wave function that describes a scalar field soared in a gravitational field of its own.

We revisit this problem on gravitating strings with much more emphasis on a description based on
ensembles of interacting random walks/polymers.\footnote{%
We give a brief comment on introduction of a statistical ensemble.
A string of a given mass $\sqrt{\mathcal{N}}/\ell_s$
is specified by superposition of level $\mathcal{N}$ states.
Since the number of states is enormous, a quantum average 
with respect to a randomly chosen state
for macroscopic observables, such as a size, density distribution and so on, agrees with a statistical average over the level $\mathcal{N}$ states, which are dominated by configurations of random walks.
Such a statistical average has also proven to be useful
to derive a (pseudo-)thermal emission spectrum
from long strings \cite{Amati:1999fv, Kawamoto:2013fza}.} %
Mathematical descriptions of random walks provide powerful tools to investigate statistical properties of polymers and so far lots of fruitful studies have been undertaken through methods such as the renormalization group \cite{Flory, DeGennes, Doi-Edwards}.
The conformation of random walks is governed by two competing tendencies;
the diffusion of monomers which is universal in systems at finite temperature and elasticity which is characteristic to long chains, both of which originate in the entropic nature of ensembles.
The balance between these effects determines the characteristic features, such as the typical shape of polymers.
Indeed, the balance between the diffusive and elastic forces results in
the typical size $R_0 \sim N^{1\over2}$ in an ideal random walk, where $N$ is the number of monomers.
A real polymer, unlike an ideal one, does not intersect with itself and is described by a self-avoiding walk (SAW).
As is well known, the self-avoiding effect makes the configuration puff up to a size of order $N^{3 \over d+2}$ in $d$ spatial dimensions \cite{Flory}.
This scaling is due to the entropic elasticity with the excluded volume effect 
taken into account.

Though fundamental strings can be described by ideal walks, it has also been suspected that when it is compressed into a small region a self-avoiding property would 
emerge nonperturbatively \cite{Susskind:1994vu}.
Therefore it is intriguing to introduce a repulsive interaction in the description of strings as in real polymers and we will describe it as a type of excluded volume effect among monomers.\footnote{%
Real polymers may be under the circumstances with various external forces like an electric potential, or interactions with solvent. We shall not argue these external effects in the present paper.} %
The other types of self-interactions are also important.
Such self-interactions include van der Waals interaction or electric repulsive interaction,
for example,
but our particular interest is gravity produced by itself.
Thus we shall investigate a long polymer with short-range repulsive 
and long-range attractive self-interactions as a model of self-gravitating strings.\footnote{%
The idea has already been mentioned in \cite{Horowitz:1997jc}, and there have been
a couple of works based on this picture \cite{KalyanaRama:1997hj, Khuri:1999ez, Khuri:1998xx}.
Our analysis partly overlaps with \cite{Khuri:1999ez}, but we here pursue more variety of scaling behavior with repulsive interaction.} %
The typical size of an interacting random walk is determined by
interplay among the repulsive and attractive self-interactions 
in addition to the diffusive and elastic forces.
Our main focus in the present paper is to see whether there exist any scaling regions in which gravity and the others are balanced, and furthermore to present an overview of the size behavior with respect to the change of the coupling constants of each self-interaction. 

This paper is organized as follows.
In the following section, we shall argue the size estimation by using 
two methods,
first by a variational method with harmonic potential, 
and second by a uniform expansion model.
The results are organized into phase diagrams in various spatial dimensions.
In section \ref{sec:conlusion-discussion}, we give a summary of the result.
Details of the calculation in the uniform expansion model and a brief argument for van der Waals interaction are presented in the appendices.

\section{Self-avoiding random walk with long range attractive force}
\label{sec:Self-repell}
In the random walk model, a long polymer is described by a chain of $N$ numbers of monomers
jointed freely with bonds.
The mathematical description for the statistical property of a self-avoiding polymer with interaction is given by Edwards Hamiltonian \cite{Edwards:1988} in $d$ spatial dimensions,
\begin{align}
  \beta H=& \frac{d}{2\ell^2} \int_0^{N} d\sigma \, \left( \frac{\partial
      \mathbf{R}}{\partial \sigma} \right)^2 
+\int_0^{N}  d\sigma \int_0^{N} d\sigma' \, V(\mathbf{R}(\sigma), \mathbf{R}(\sigma'))
\label{eq:Edwards_H}
\,.
\end{align}
Here the potential term consists of long-range Newton interaction
in $d \,(>2)$ dimensions as well as a point-like repelling force expressed by
the delta function,
\begin{align}
  V(\mathbf{R}(\sigma), \mathbf{R}(\sigma'))
=& -\frac{g^2 \ell^{d-2}}{|\mathbf{R}(\sigma)- \mathbf{R}(\sigma')|^{d-2}}
+u \ell^{d} \delta^{(d)}(\mathbf{R}(\sigma)- \mathbf{R}(\sigma')) \,,
\label{eq:V_explicit}
\end{align}
where $g^2$ and $u$ are dimensionless coupling constants.
$\ell$ is the (Kuhn) length of the bond between monomers,
 which will be
identified with the string scale $\ell_s$.
We are considering an ensemble of highly excited long strings of level $\mathcal{N}$,
and the corresponding temperature $\beta^{-1}$ is on the order of the string scale $1/\ell_s$,
but its explicit value is not relevant to our analysis.
This sets the  scale of the analysis, and the coupling constants are measured with respect
to this scale.
Since the length along a string is proportional to $\ell_s \sqrt{\mathcal{N}}$,
the excited level is related to the number of monomers as $N \propto \sqrt{\mathcal{N}}$.
The Hamiltonian \eqref{eq:Edwards_H} can be understood as a continuum version of a discrete
model (a bead-spring model with interaction),
\begin{align}
  \beta H =& \frac{d}{2\ell^2 } \sum_{n=1}^N |\mathbf{R}_n - \mathbf{R}_{n-1}|^2
+ \sum_{n\neq n'} V(\mathbf{R}_{n}, \mathbf{R}_{n'}) \,,
\label{eq:Edwards_H_disc}
\end{align}
where $\mathbf{R}_n$ is the position vector of $n$-th monomer.
In the interaction term, $n=n'$ terms are excluded since they correspond to
the self-interaction of each monomer.\footnote{%
Precisely speaking, 
a free part should have been written as
$\frac{d}{2b^2} \sum_{n=1}^{\hat{N}} |\mathbf{R}_n - \mathbf{R}_{n-1}|^2$ 
with a fundamental bond length $b$ and a total number of monomers $\hat{N}$.
The continuum limit is taken by
$b \rightarrow 0$ and $\hat{N} \rightarrow \infty$
with the total ``diffusion time'' $T= \hat{N} b^2$ fixed, and
the parametrization on the polymer is $t=n b^2$ ($0\leq t \leq T$).
 (This is the continuum limit
for the diffusion equation that a random walk probability distribution satisfies.)
Kuhn length $\ell$ and an effective number of monomers $N$ are introduced so that
the size, the total mass and length be $\ell N^{1\over2}$, $N/\ell$ and $N\ell$ respectively;
namely a polymer is effectively regarded as $N$ number of monomers (of unit mass)
joined by bonds of length $\ell$. $T$ is identified with $N\ell^2$, and
a continuum dimensionless variable $\sigma=t/ \ell^2$ 
 ($0 \leq \sigma  \leq N$) is introduced to represent a position on the polymer.}
In the continuum version \eqref{eq:Edwards_H}, this self-interaction of each single monomer
should also be excluded and we can understand the coupling constants are suitably
renormalized ones (with appropriate regularization such as point-splitting).

What we want to understand is how the size varies as
the couplings $g^2$ and $u$ change.
In particular, we shall explore if it exhibits any scaling behavior
with respect to the number of the monomers $N$.
Our primal goal is to evaluate
the end-to-end radius squared,
\begin{align}
  \vev{\mathbf{R}^2} =&
\frac{1}{Z} \int \mathcal{D}\mathbf{R}(N)  \,
\big( \mathbf{R}(N) - \mathbf{R}(0) \big)^2 \,
e^{-\beta H} \,,
\qquad
Z= \int \mathcal{D}\mathbf{R} \,
e^{-\beta H} 
\,,
\end{align}
where the end point $\mathbf{R}(N)$ is integrated over while
the other end is fixed due to the translational invariance.
Hereafter, $\mathbf{R}(0)$ is fixed at the origin.

Before starting the analysis, we roughly estimate the $N$ dependence
of the coupling constants with which the typical size of long polymers gets affected.
As is well-known, with no interaction $g^2=u=0$, a typical size of free long polymers is
given by $R_0 \simeq \ell N^{1\over2}$ (hereafter $R_0$ always stands for this free walk size).
When the coupling constants are turned on, the size will be altered but the $N$ dependence remains the same until the couplings reach certain marginal values.
Such marginal values, $g_o$ and $u_o$,
are given by the condition that the interaction terms are of $O(1)$.
Since the distance of two different monomers scales as $|\mathbf{R}(\sigma)-\mathbf{R}(\sigma')|
\sim N^{1\over2}$,
the gravitational force becomes significant at $g_o$, where
$N^2 g^2_o N^{-\frac{d-2}{2}} \sim O(1)$ and the first $N^2$ comes from the number
of pairs.
This condition gives $g_o \sim N^{\frac{d-6}{4}}$ \cite{Horowitz:1997jc, Khuri:1999ez}.
On the other hand, the repulsive interaction is local.
The volume occupied by the polymer is proportional to $N^{d\over2}$, 
and then the chance of two different monomers 
coming to the same point is $N^{-{d\over2}}$.
Hence, the marginal coupling is determined by 
$N^2 u_o N^{-{d\over2}} \sim O(1)$, or $u_o \sim N^{\frac{d-4}{2}}$.
This result suggests that for $d>4$ the contact interactions 
become negligible, and this is consistent with the well-known fact 
that the probability of self-intersection of free random walks is negligible for $d>4$.
So far, we have investigated the marginal couplings 
at which each force takes effect against the entropic elasticity.
It is possible that the gravitational and the repulsive forces 
balance out to sustain a configuration.
This is realized when 
$g_o^{\prime 2} N^{-\frac{d-6}{2}} \sim u_o' N^{-\frac{d-4}{2}}$
(the prime is put to indicate that a different type of equilibrium is realized), or $N g_o^{\prime 2} /u_o' \sim O(1)$.
We will see that the following calculation reproduces this condition.

\subsection{Evaluation of size by variational method with harmonic potential}
\label{sec:harmonic}

We shall evaluate the averaged size-squared $\vev{\mathbf{R}^2}$ by the variational principle.
The trial Hamiltonian is chosen to be a harmonic action,
\begin{align}
  \beta H_0 =&
\frac{d}{2 \ell^2}
\int_0^{N} d\sigma \, \left( \frac{\partial \mathbf{R}}{\partial \sigma} \right)^2
+\frac{d q^2}{2\ell^2} \int_0^N d\sigma \, \mathbf{R}(\sigma)^2 \,,
\end{align}
where $q$ is a dimensionless variation parameter.\footnote{%
This corresponds
to the one used in \cite{Edwards:1988} as
$q_\text{ours} = \frac{\ell}{3} q_\text{EM}$.}
The free energy,
\begin{align}
    e^{-\beta F}=&
\int \mathcal{D}\mathbf{R} \,
e^{-\beta H_0 - \beta(H-H_0)} \,,
\nn\\
\beta(H-H_0) =&
\int_0^{N} d\sigma  \int_0^{N} d\sigma'
\, V(\mathbf{R}(\sigma), \mathbf{R}(\sigma'))
-\frac{d q^2}{2\ell^2} \int_0^N d\sigma \, \mathbf{R}(\sigma)^2 \,,
\end{align}
satisfies the following inequality,
\begin{align}
\beta F \leq 
\beta {F}_0(q) + \beta\vev{H-H_0}_0
\,,
\label{eq:variational_free_energy}
\end{align}
where $\vev{\cdots}_0$ is the expectation value with respect to the trial Hamiltonian $\beta H_0$ and $\beta F_0$ is the corresponding free energy.
The variation parameter $q$ will be chosen so that it minimizes the right hand side of the inequality, and we shall estimate averaged sizes by using the optimized parameter.

The trial Hamiltonian provides the propagator
\begin{align}
\label{eq:harmonic_propagator_full}
    G(\sigma,\sigma') =&
\left(
\frac{qd}{2\pi\ell^2 \sinh {q|\sigma - \sigma'|}}
\right)^{\frac{d}{2}}
\exp\left(
-\frac{qd
\left[
\big( \mathbf{R}(\sigma)^2 + \mathbf{R}(\sigma')^2 \big)
 \cosh {q|\sigma - \sigma'|}
-2 \mathbf{R}(\sigma) \cdot \mathbf{R}(\sigma')
\right]
}{2\ell^2 \sinh {q|\sigma - \sigma'|}}
\right)
\,.
\end{align}
The expectation value of a function of $k$ different points $\mathbf{R}(\sigma_i)$
with respect to $\beta H_0$ is calculated by use of
$G(\sigma,\sigma')$ as
\bea
\vev{\mathcal{O}\big( \mathbf{R}(\sigma), \cdots, \mathbf{R}(\sigma_k) \big) }_0
&=&
\frac{\int \mathcal{O} e^{-\beta H_0}}{\int e^{-\beta H_0}}
\nn \\
&=&
\frac{1}{Z_0}
\int \prod_{i=1}^{k+1} \big[ \mathcal{D}\mathbf{R}(\sigma_i) G(\sigma_{i-1},\sigma_i) \big]
\,\mathcal{O} \big( \mathbf{R}(\sigma), \cdots, \mathbf{R}(\sigma_k) \big)
\label{def:vev_by_prop}
 \,,
\eea
with $\sigma_0=0$, $\sigma_{k+1}=N$,
and
the partition function is
\begin{align}
Z_0 =  \int \mathcal{D}\mathbf{R}(N) \,  G(0,N)
= \cosh^{-\frac{d}{2}} {q N} \,.
\end{align}
The expectation value of the size-squared with respect to $\beta H_0$ is 
\begin{align}
    \vev{\mathbf{R}^2}_0 =&
\frac{1}{Z_0}\int \mathcal{D}\mathbf{R}(N) \, \mathbf{R}(N)^2 G(0,N)
=\frac{\ell^2}{q} \tanh {q N} \,.
\label{eq:2}
\end{align}

In the following, we shall find optimized parameters, 
which we call $q_0$, 
that minimize the free energy bound \eqref{eq:variational_free_energy}.
They will be given in terms of $u$ and $g$, and determine the size behavior in the space of the coupling constants.
We claim that the mean radius of configurations $\vev{\mathbf{R}^2}$, which we shall denote $R^2$, is approximated by using the optimized parameter as
\bea
R^2 \simeq \vev{\mathbf{R}^2}_0|_{q=q_0}
=\frac{\ell^2}{q_0} \tanh {q_0 N} \,.
\label{eq:approx_size}
\eea
From this expression, one can see that the size stays to be the free walk one $R_0$
for $q_0 N \ll 1$ and it changes from $q_0 N \sim O(1)$ as
\begin{align}
  R \simeq &
             \begin{cases}
               \ell \sqrt{N} & q_0 N \ll 1 \\[1em]
               \dfrac{\ell}{\sqrt{q_0}} & q_0 N \gtrsim O(1)
             \end{cases} \,.
\label{eq:size_scaling_q0}
\end{align}

Now we evaluate the term $\vev{\beta(H-H_0)}_0$
in the right hand side of \eqref{eq:variational_free_energy}.
For the quadratic potential part, we find 
\begin{align}
\frac{qd^2}{2\ell^2} \int_0^N d\sigma \,
\vev{\mathbf{R}(\sigma)^2}_0
=
\frac{q}{2} \frac{\partial (\beta F_0)}{\partial q}
= \frac{qd\, N}{4}
\tanh {q N}
 \,.
\end{align}
As for the original potential term, we first note that 
the part including the Newton potential can be rewritten in terms of an exponential,
\begin{align}
  \vev{\frac{1}{|\mathbf{R}(\sigma) - \mathbf{R}(\sigma')|^{d-2}}}_0
=
\frac{i^{\frac{d-2}{2}}}{\Gamma(\frac{d-2}{2})}
\int_0^\infty dx x^{\frac{d}{2}-2} 
\vev{e^{-ix (\mathbf{R}(\sigma) - \mathbf{R}(\sigma'))^2-\epsilon x}}_0 \,,
\label{eq:Coulomb_exp}
\end{align}
where $\epsilon>0$ is a regulator.
The integrals with respect to $\mathbf{R}(\sigma)$ are just Gaussian. Therefore a straightforward calculation gives the expectation value\footnote{
In what follows, we take $\sigma \leq \sigma'$ by
writing the integral as
$\int_0^N d\sigma \int_0^N d\sigma' = 2 \int_0^N d\sigma' \int_0^{\sigma'} d\sigma$, since the integrand is symmetric under the interchange of  $\sigma$ and $\sigma'$.
Then the results do not appear 
in a symmetric way for exchanging $\sigma$ and $\sigma'$.}
\bea
&&\vev{\frac{1}{|\mathbf{R}(\sigma) - \mathbf{R}(\sigma')|^{d-2}}}_0
=\frac{1}{\Gamma(\frac{d}{2})}
\left(\frac{qd}{2\ell^2 F_1(\sigma,\sigma';q)} \right)^{\frac{d-2}{2}}
\,,
\eea
where
\begin{align}
F_1(\sigma,\sigma';q)
=&
\frac{\sinh q \sigma \cosh q(N-\sigma) + \sinh q \sigma' \cosh q(N-\sigma')-2\sinh q\sigma \cosh q(N-\sigma')}{\cosh q N}
\,.
\label{eq:F1}
\end{align}

For the repulsive interaction part, we rewrite the expectation value of the delta function in the integral form.
The integrals are again Gaussian, and we easily find
\bea
\vev{\delta^{(d)}(\mathbf{R}(\sigma) - \mathbf{R}(\sigma'))}_0
&=&
\int \frac{d^d k}{(2\pi)^d}
\vev{e^{i \mathbf{k} \cdot [ \mathbf{R}(\sigma) - \mathbf{R}(\sigma')]}}_0
=
\left( \frac{qd}{2\pi \ell^2 F_2(\sigma,\sigma';q)} \right)^{\frac{d}{2}}
\,,
\eea
where
\begin{align}
F_2(\sigma,\sigma';q)=&
\frac{\sinh q \sigma \sinh q(\sigma'-\sigma)}{\sinh q\sigma'}
+\frac{\cosh q(N-\sigma') \sinh q\sigma'}{\cosh qN}
\bigg(1- \frac{\sinh q\sigma }{\sinh q\sigma'} \bigg)^2  
\,.
\end{align}

Combining these into \eqref{eq:variational_free_energy}, we find the inequality
\bea
  \label{eq:free_energy_harmonic_q}
  \beta F & \leq &
\frac{d}{2} \ln \big( \cosh qN \big) 
- \frac{qdN}{4} \tanh qN
\nn \\
&& - 2 \int_0^N d\sigma' \int_0^{\sigma'} d\sigma
\,
\left[
\frac{g^2}{\Gamma(\frac{d}{2})}\left({qd \over 2F_1(\sigma,\sigma';q)} \right)^{\frac{d-2}{2}} 
-u \left({qd\over 2\pi F_2(\sigma,\sigma';q)}\right)^{\frac{d}{2}}
\right]
\,.
\eea
We need to tune $q$ to find the most strict bound for the free energy.
It is, however, difficult to analytically evaluate $\sigma$ and $\sigma'$ integrals, since
$F_1(\sigma,\sigma';q)$ and $F_2(\sigma,\sigma';q)$
are complicated functions of
$\sigma$ and $\sigma'$.
We instead use an approximation to proceed.
We take all the exponential quantities, namely $e^{-q\sigma}, e^{-q\sigma'}, e^{-qN}, e^{-q(\sigma'-\sigma)}, e^{-q(N-\sigma)}$ and $e^{-q(N-\sigma')}$, 
to be small and negligible.\footnote{%
With this approximation the propagator \eqref{eq:harmonic_propagator_full} becomes
that of the ground state approximation,
\begin{align}
        G_0(\sigma,\sigma') \simeq &
\left( \frac{dq}{\pi \ell^2 }\right)^{\frac{d}{2}}
\exp \bigg[
-\frac{qd}{2\ell^2} \big( \mathbf{R}(\sigma)^2 + \mathbf{R}(\sigma')^2 \big)
-\frac{qd}{2} |\sigma-\sigma'|
\bigg] \,.\nn
\end{align}
Thus, this can be viewed as the application of the ground state approximation to
the evaluation of the right hand side of \eqref{eq:free_energy_harmonic_q},
which is valid if the dimensionless level separation $q$ is not much small.}
This will not hold near the ends and the point at which $\sigma=\sigma'$ 
in the integration region.
These points are in reality separated by a fundamental bond length,
and the approximation will be valid elsewhere as long as $q$ is not
small enough.
With this approximation, the free energy bound is simplified to be
\bea
\beta F &\leq &
\frac{qdN}{4}
-N^2 \left[ {g^2 \over \Gamma(\frac{d}{2})}
\left({qd \over 2}\right)^{\frac{d-2}{2}} 
- u \left({qd \over 2 \pi} \right)^{\frac{d}{2}} \right]
\,.
\label{eq:F_harmonic_0}
\eea
With unimportant positive numerical factors omitted, the extremal condition becomes
\bea
1 -N g^2 q_0^{\frac{d-4}{2}}+N u q_0^{\frac{d-2}{2}} = 0 \,.
\label{eq:extrem_eq_gu}
\eea
In the following, we examine \eqref{eq:F_harmonic_0} and \eqref{eq:extrem_eq_gu}
to find an optimal value $q_0$.
We start with a pure gravitational case ($g>0, u=0$) to see the method rederives the
already known behavior in \cite{Horowitz:1997jc, Khuri:1999ez}.
Next, a pure repulsive case ($u>0, g=0$) is argued and a limitation of the current
variational method is presented.
Finally, we consider a generic case ($g,u>0$) in various dimensions and discuss
that there appear two different size scalings with respect to the magnitude of $u$.

We first consider the case with no repulsive force, namely $u=0$.
This corresponds to the original self-gravitating fundamental string studied in \cite{Horowitz:1997jc, Khuri:1999ez}.
The results depend on the spatial dimension $d$. 
For $d$ other than 4 (for which a separate argument will be given shortly), the extremal value is $ q_0 \simeq  \big(g^4 N^2 \big)^{-\frac{1}{d-4}}$ (a numerical coefficient is of no importance).
In $d<4$, the right hand side of \eqref{eq:F_harmonic_0} is a convex function of $q$
when $g^2>0$, and then $q_0$ indeed provides the optimal value (an absolute minimum),
and the marginal coupling $g_o$ is obtained form the relation $q_0 N \simeq O(1)$ as $g_o \simeq N^{d-6 \over 4}$ which is consistent with the previous estimation at the beginning of this section.
The size behavior is easily read off from
\begin{align}
R
\simeq
  \begin{cases}
    \ell \sqrt{N} & g < g_o  \\
    \ell \big( g^2 N \big)^{\frac{1}{d-4}}  & g \geq g_o
  \end{cases}
\label{eq:size_harmonic_puregrav_gen}
\,.
\end{align}
The size would change as the coupling grows and becomes comparable with the Schwarzschild radius $R_s \simeq \ell(g^2N)^{1 \over d-2}$ at the critical coupling $g_c \sim N^{-{1\over2}}$, where we have identified $\ell$ with the string fundamental length $\ell_s$.

In $d=4$, the free energy satisfies the inequality
$\beta F \lesssim  q {N} \big(1 - g^2 {N} \big)$.
Note that the marginal and the critical couplings are the same order, 
namely $g_o \sim g_c\sim N^{-{1\over2}}$. 
When $g \lesssim g_o$, the optimal value is $q_0=0$,
which gives the free random walk behavior.
On the other hand, when $g \gtrsim g_o$, we find $q_0 \rightarrow \infty$ and $R \rightarrow 0$ as is obvious from \eqref{eq:size_scaling_q0}.
This means that the polymer suddenly collapses at $g_o$.
In the case $d>4$, the marginal coupling $g_o$ is larger than $g_c$,
and the system is suspected to exhibit a hysteresis \cite{Horowitz:1997jc}.
Since the right hand side of \eqref{eq:F_harmonic_0} is a concave function, the extremal value $q_0 \simeq  \big(g^4 N^2 \big)^{-\frac{1}{d-4}}$ leads to a local maximum, and
the correct optimal value of $q$ is $q_0=0$ for $g^2=0$ and $q_0=\infty$ for $g^2>0$.
This implies that the configuration completely collapses from a free walk size $R_0$ with an arbitrary small value of $g$.
The behavior is more exotic than discussed in  \cite{Horowitz:1997jc}
(this is also mentioned in \cite{Khuri:1999ez}).
In this pure attractive case (especially $2<d\leq 4$),  
we have rederived the results obtained by
Horowitz-Polchinski \cite{Horowitz:1997jc} and Khuri \cite{Khuri:1999ez}.

Next, we examine the case with a repulsive interaction and take $u \neq 0$.
To begin with, we give a comment on the case with a pure repulsive force $u>0$, $g^2=0$.
This should correspond to a SAW without the attractive interaction,
and one may expect to obtain the famous result by Flory \cite{Flory},
$R \sim N^{\frac{3}{d+2}}$ ($1 \leq d \leq 4$).
From \eqref{eq:F_harmonic_0} one can immediately see
that the only solution is $q_0=0$ when $g^2=0$ since all the coefficients
are positive.\footnote{%
Note that $q_0$ has to be non-negative
otherwise the expectation values are not well-defined.}
This provides the free walk behavior of the size
$R_0$ and fails to reproduce Flory's result.
Indeed, as long as we consider a confining harmonic potential,
the size of the walk should decrease compared to the
free walk result for any choice of $q$.
One sees $q_0=0$ stands for no harmonic potential, so the configuration expands
as a free walk, but it cannot spread larger than that size.\footnote{%
This argument can also be applied to the situation with $g^2$ sign flipped,
$g^2 \rightarrow -g^2$.
Together with a repulsive force $u>0$, this describes a single polyelectrolytes chain
in an ideal situation where no screening effect from solvent takes place.
By the Flory's scaling argument the size behavior is evaluated to be 
$R \sim N^{\frac{3}{d}}$ \cite{Pfeuty}, and furthermore by employing a renormalization group argument $R \sim N^{\frac{2}{d-2}}$ \cite{DeGennes, Pfeuty};
they are much larger than the free walks, 
but our calculation merely gives a free walk behavior.}
Based on this observation, we conclude that the variational calculation with a harmonic potential 
is not suitable for describing  random walks that might expand larger than the free walk size $R_0$.
From \eqref{eq:size_scaling_q0}, one sees that the size 
reaches to the fundamental length scale at $q_0 \sim O(1)$, and the description based on this effective Hamiltonian may not be valid after that point, providing the upper bound on $q_0$.
Thus, the validity of this variational method will be as follows:
if the maximal size of a configuration is known, a priori, to be $R_0$, we may trust
the method for $q_0 \lesssim O(1)$, but if it is possible that a configuration can expand
larger than $R_0$, it is safe to take $q_0$ to be $N^{-1} \lesssim q_0 \lesssim O(1)$.
At any rate, this variational method based on the harmonic potential will serve a reasonable
size evaluation if the size is smaller than or equal to $R_0$.

Now we turn on the gravitational force, $g^2 > 0$.
The only negative term in \eqref{eq:extrem_eq_gu}, $-N g^2 q_0^{\frac{d-4}{2}}$,
will balance out with the dominant one between the positive terms, $1$ and $N u q_0^{\frac{d-2}{2}}$
(or both if they are comparable).
If the gravitational force term balances with the first term, $1$, the solution
is $q_0 \simeq  \big(g^4 N^2 \big)^{-\frac{1}{d-4}}$ as in the previous pure gravity case
(we give a separate consideration for the $d=4$ case later).
This solution is valid as long as $N u q_0^{\frac{d-2}{2}} \ll 1$ for a given value of
$q_0$.
This condition is rephrased in terms of a condition for $g$,
\begin{align}
  g < & \tilde{g}_o \quad (2<d<4) \,,
\end{align}
where
\begin{align}
\tilde{g}_o \simeq u^{\frac{d-4}{2(d-2)}} N^{-\frac{1}{d-2}} \,. 
\end{align}
In $d>4$, the condition for this $q_0$ to be valid is $g> \tilde{g}_o$,
but
the configuration with pure gravity (namely, without the repulsive force) exhibits
a strange behavior as we have seen, and the variational method itself does not
seem valid.
On the other hand,
if the third term in \eqref{eq:extrem_eq_gu} is dominant over the first term, the gravitational and repulsive forces balance, and the solution is $q_0 \simeq g^2/u$.
The consistency condition, $N u q_0^{\frac{d-2}{2}} \gg 1$, implies
that $g <  \tilde{g}_o$ for $d>2$.
Thus, $\tilde{g}_o$ is a marginal coupling at which the solution of the stationary
condition switches from $q_0 \simeq  \big(g^4 N^2 \big)^{-\frac{1}{d-4}}$ to
$q_0 \simeq g^2/u$.
At $g=\tilde{g}_o$, all the terms in \eqref{eq:extrem_eq_gu} are of the same order,
and these two expressions of $q_0$ agree.
The observation so far can be demonstrated by use of an explicit solution of $q_0$
in $d=3$,
\begin{align}
{q}_0=& \bigg[\frac{1}{2N {u}}\big( -1+\sqrt{1+4N^2{g}^2{u}} \big)  \bigg]^2 \,.
\end{align}
In $d=3$, $\tilde{g}_o \simeq u^{-\frac{1}{2}} N^{-1}$.
Thus, $g< \tilde{g}_o$ (in terms of $N$ dependence)
implies that $N^2 g^2 u \ll 1$, and in this region $q_0$ is reduced to
$q_0 \simeq g^4 N^2$.
On the other hand, when $g>\tilde{g}_o$, we have $q_0 \simeq g^2/u$.
Around $g \simeq \tilde{g}_o$ ($N^2 g^2 u \sim O(1)$),
 the entropic elastic, the gravitational, and the repulsive forces balance out,
and $q_0 \simeq (Nu)^{-2}$, which can be expressed as $g^4 N^2$ or $g^2/u$ by use of
$u \simeq (gN)^{-2}$.

We move on to the separate case in $d=4$,
where $\tilde{g}_o \sim N^{-\frac{1}{2}}$.
We find
\begin{align}
{q}_0=& \frac{g^2N-1}{uN} \,.
\end{align}
Note that $g^2 N$ has to be greater than or equal to $1$ since $q_0$ must be non-negative.
For $g< \tilde{g}_o$, namely $g^2 N < 1$, $g^2$ term in \eqref{eq:extrem_eq_gu}
is $q$ independent and subleading.
We then come back to \eqref{eq:F_harmonic_0}
and find that $q_0=0$.
For $g > \tilde{g}_o$, $q_0$ becomes $g^2/u$, and
for $g \simeq \tilde{g}_o$ ($g^2N \sim O(1)$),
$q_0 \simeq (uN)^{-1} \simeq g^2/u$.
Thus, in $d=4$, the optimal value of $q_0$ changes from $q_0=0$ to
$q_0=g^2/u$ at $g=\tilde{g}_o$.

Let us estimate the size scaling in the case of $q_0=g^2/u$ in arbitrary $d>2$.
The mean square size is given by \eqref{eq:approx_size},
\begin{align}
R^2
\simeq \frac{\ell^2 u}{g^2} \tanh \bigg(\frac{g^2 N}{u} \bigg)
\simeq
  \begin{cases}
    \ell^2 N & g^2 N/u \ll 1  \\
    \ell^2 u/g^2  & g^2 N/u \gg 1
  \end{cases}
\label{eq:ms_size_harmonic_gen}
\,.
\end{align}
Thus, when the repulsive force is much stronger than the attractive force, $u \gg g^2 N$,
the size is given by just the free random walk one, but does not expand larger than that.
The feature that the size does not exceed the free walk size 
would be a limitation of this approximation as we have discussed in the pure repulsive case.
In the opposite limit, the behavior is affected by $N$ dependence of the interactions.
Here we consider the $N$ dependence of $u$ to be fixed, and observe the scaling of the size by changing $g^2$.
When $q_0 N \sim O(1)$, the size starts to change and this defines the marginal value
\begin{align}
  g_o' \simeq \sqrt{\frac{u}{N}} \,,
\end{align}
which agrees with the naive scaling argument at the beginning of this section.
As the coupling grows, the size decreases, and it eventually coincides with the Schwarzschild radius of a black hole
of the same total mass; $R_s \simeq \ell (g^2 N)^{\frac{1}{d-2}}$
in $d$ dimensions.
The value of the critical coupling at which this transition takes place
is
\begin{align}
  g_c' \simeq
u^{\frac{d-2}{2d}} N^{-\frac{1}{d}} \,,
\label{eq:critical_coupling}
\end{align}
and the size of the corresponding black hole is
\begin{align}
  R_c \simeq \ell (uN)^{\frac{1}{d}} \,.
\label{eq:critical_radius}
\end{align}

Let us summarize the size scaling in this generic case.
The size depends on the $N$-dependence of the coupling constants
$g$ and $u$.
Here, we observe the change of the size by tuning
the gravitational coupling $g$ from $g=0$ to larger values
for a given value of $u$, until the size of the configuration becomes
the Schwarzschild radius of a black hole of the same mass.
For a small value of $g < \tilde{g}_o$, the first two terms in \eqref{eq:extrem_eq_gu}
balance, and the size scaling is the one without a repulsive force
given in \eqref{eq:size_harmonic_puregrav_gen}.
Thus, for sufficiently small couplings $g< \tilde{g}_o, g_o$,
the size is  the free walk one $R_0$.
If the repulsive force coupling is larger than the marginal value $u_o \sim N^{\frac{d-4}{2}}$,
the configuration is expected to be expanded larger than $R_0$, but the variational calculation
is not capable of realizing such an expanded configuration as argued.
In order to obtain a consistent picture,
we take $u <  u_o$ with which the size is to be $R_0$ for a very small $g < \tilde{g}_o, g_o$,
and leave the analysis for $u> u_o$ with small $g$ to the next subsection where another approximation method
is introduced.
For larger values of $g$,
the behavior depends on the magnitude relation among
$g_o$, $\tilde{g}_o$, and $g_o'$.
It is not difficult to check that $g_o' < \tilde{g}_o$ if $u < u_o$.
Thus, in this region of $u$, the configuration immediately starts to shrink
as $R \simeq \ell \sqrt{u}/g$ 
once the repulsive force participates in the balance
at $g = \tilde{g}_o$.
If $g_o < \tilde{g}_o$, the size shrinks as $R \simeq \ell (g^2 N)^{\frac{1}{d-4}}$
from $g=g_o$ and changes its behavior to $R \simeq \ell \sqrt{u}/g$ at $g = \tilde{g}_o$,
and the configuration will eventually be covered by the Schwarzschild radius $R_c$ at $g=g_c'$.
However, if $u$ is too small, the configuration becomes a black hole
before the repulsive force becomes important. 
It happens when $g_c < \tilde{g}_o$, which leads to $u < N^{-1}$,
and the configuration collapses to a black hole of radius $\ell$ at $g=g_c$.\footnote{%
From \eqref{eq:size_scaling_q0}, one finds that the size becomes comparable to
a fundamental scale $\ell$ at $q_0 \sim O(1)$.
If $u< N^{-1}$, the $u$ term in \eqref{eq:extrem_eq_gu} remains subleading
for $q_0 < O(1)$, and then negligible in the whole process.}
If $\tilde{g}_o < g_o$, the size starts to change as $R \simeq \ell \sqrt{u}/g$ 
at $g = \tilde{g}_o$ from the beginning, and it becomes a black hole at $g=g_c'$.

To illustrate a typical size behavior, we pick up a couple of examples in $d=3$.
If $u$ takes the marginal value $u_o\sim N^{-{1\over2}}$,
the repulsive force pushes the critical coupling to a larger value $g_c' \sim N^{-{5\over 12}}$
than that of non repulsive case, $g_c\sim N^{-{1\over2}}$,
and the black hole swells ($R_c \simeq \ell N^{1\over 6}$).
As we will see in Sec.~\ref{sec:uniform}, real polymers have $u \sim O(1)$,
which corresponds to a well-known phenomenological size scaling by Flory.
If $ u \sim O(1)$ in $d=3$, the critical coupling and the size at the corresponding point are
even larger,
$g_c' \sim N^{-{1\over 3}}$ and $R_c \sim \ell N^{1\over 3}$ respectively.\footnote{%
As noted a couple of times so far, when $u>u_o$,
the variational method does not give a reasonable
size scaling as long as $g$ is small,
but the critical coupling and size here are indeed valid
as they agree with the result by another method we are about to introduce.}
Note that from \eqref{eq:critical_radius}
one can see that
$R_c$ corresponds to a close-packed configuration of $N$ monomers in $d$ dimensions for $u \sim O(1)$.
In $d=4$, $g_o \simeq \tilde{g}_o$, and one can see that the size suddenly changes
from $R_0$ to $R \simeq \ell \sqrt{u}/g$ at $g=g_o \sim N^{-\frac{1}{2}}$.
Especially, if $u< N^{-1}$, the configuration collapses in to a small black hole
of size $\ell$.
More detailed observation on the size and the values of the critical couplings are
presented in Sec.~\ref{sec:phase-diagram} where
phase diagrams are drawn.

So far we have obtained the size scalings which
are reliable for the configuration which shrinks due to the attractive force.
In order to investigate configuration larger than $R_0$,
we need to invoke some other methods.
In the following subsection, we shall consider a different approximation scheme that turns out to work for expanded configurations.

\subsection{Evaluation of size by uniform expansion model}
\label{sec:uniform}

In this subsection, we try another approximation to evaluate the change of the size
of self-gravitating polymers.
This method is called the uniform expansion model (UEM)
\cite{Doi-Edwards}, in which the fundamental length of the bond $\ell$ is renormalized as 
$\ell'=a \ell$ ($a>0$) but the configuration itself
is assumed to remain that of the free random walk.

We consider an effective free Edwards Hamiltonian with the bond length $\ell'=a \ell$,
\begin{align}
  \beta H' =&
 \frac{d}{2a^2\ell^2} \int_0^{{N}}d{\sigma}
\bigg( \frac{\partial \mathbf{R}}{\partial {\sigma}} \bigg)^2
\,.
\label{eq:Edwards_UEM}
\end{align}
The Green function corresponding to this Gaussian action
is
\begin{align}
  G'(\sigma,\sigma')=&
\left(
\frac{d}{2\pi \ell^2 a^2 |\sigma-\sigma'|}
\right)^{\frac{d}{2}}
\exp \bigg[-\frac{d}{2\ell^2 a^2|\sigma-\sigma'|} 
\big(\mathbf{R}(\sigma)-\mathbf{R}(\sigma') \big)^2 \bigg]
\,,
\end{align}
and the expectation value with respect to $\beta H'$,
$\vev{\cdots}'$,
 is calculated with this Gaussian propagator in the same way as
\eqref{def:vev_by_prop}, where the partition function is unity.

The end-to-end distance squared, evaluated with respect to the original action including the interactions \eqref{eq:Edwards_H}, is rewritten as an expectation value
with respect to $\beta H'$ as
\begin{align}
  \vev{\mathbf{R}^2} =&
\frac{\int \big( \mathbf{R}(N)-\mathbf{R}(0) \big)^2 e^{-\beta H}}{\int e^{-\beta H}}
= \frac{\vev{\big( \mathbf{R}(N)-\mathbf{R}(0) \big)^2e^{-\beta(H-H')}}'}{\vev{e^{-\beta(H-H')}}'}
\,.
\end{align}
In UEM,  we assume that $\beta(H-H')$ can be treated as a perturbation,
and then
\begin{align}
  \label{eq:94}
  \vev{\mathbf{R}^2} \simeq &
\frac{\vev{(\mathbf{R}(N)-\mathbf{R}(0))^2(1-\beta (H-H')
    )}'}{\vev{1-\beta(H-H')}'}
\nn\\ \simeq &
\vev{(\mathbf{R}(N)-\mathbf{R}(0))^2}' \big(1+ \vev{\beta(H-H')}' \big)
-\vev{(\mathbf{R}(N)-\mathbf{R}(0))^2 \beta(H-H')}'
\,,
\end{align}
to the first order. 
One can easily see $\vev{(\mathbf{R}(N)-\mathbf{R}(0))^2}' = a^2 \ell^2 N$, since it is the radius-squared of a free random walk with $\ell'$.
The other terms are spelled out as
\bea
\vev{\beta(H-H')}'
=
\frac{d}{2\ell^2}\bigg(1-\frac{1}{a^2} \bigg)
\int_0^N d\sigma \, 
\vev{ \left( \frac{\partial \mathbf{R}}{\partial \sigma}\right)^2}'
+ \int_0^N d\sigma \int_0^N d\sigma'
\vev{V(\sigma,\sigma')}' \,,
\label{eq:UEM_VEV1}
\eea
and
\bea
\vev{(\mathbf{R}(N)-\mathbf{R}(0))^2\beta(H-H')}'
&=&
\frac{d}{2\ell^2}\bigg(1-\frac{1}{a^2} \bigg)
\int_0^N d\sigma \, 
\vev{ \left( \frac{\partial \mathbf{R}}{\partial \sigma}\right)^2
 (\mathbf{R}(N)-\mathbf{R}(0))^2}'
\nn\\
&&
+ \int_0^N d\sigma \int_0^N d\sigma'
\vev{V(\sigma,\sigma')\, (\mathbf{R}(N)-\mathbf{R}(0))^2}' \,.
\label{eq:UEM_VEV2}
\eea
We evaluate these expectation values in Appendix \ref{sec:misc-calc}
in which the details are presented.
The result of the expectation value of the size squared is summarized as
\begin{align}
  \vev{\mathbf{R}^2}=&
N \ell^{\prime 2} +Na^2 \ell^2 \big( 1-a^2 \big)
+2 u \ell^2 a^2
\left(
\frac{d}{2\pi a^2}
\right)^{\frac{d}{2}}
 \int_0^N d\sigma' \int_0^{\sigma'} d\sigma \,
(\sigma'-\sigma)^{\frac{2-d}{2}} 
\nn\\&
-
\frac{2 g^2 \ell^{2} a^2}{\Gamma(\frac{d}{2})}
\bigg(\frac{d}{2a^2}\bigg)^{\frac{d-2}{2}}
\frac{d-2}{d}
\int_0^N d\sigma' \int_0^{\sigma'}d\sigma \,
 (\sigma'-\sigma)^{\frac{4-d}{2}} \,.
\label{eq:R2_UEM}
\end{align}
The integrals should be understood as the double sum, $\sum_{n'=2}^N \sum_{n=1}^{n'-1}$,
but it is much easier to evaluate them by using continuum variables $\sigma$
and $\sigma'$.
The result becomes
\begin{align}
  \label{eq:108}
  \vev{\mathbf{R}^2}=&
\left[
N a^2 
+Na^2  \big( 1-a^2 \big)
+C_1 u N^{\frac{6-d}{2}} a^{2-d} 
-C_2 g^2 N^{\frac{8-d}{2}} a^{4-d} 
\right]\ell^2
\,,  
\end{align}
where
\begin{align}
  \label{eq:117}
  C_1=2 \bigg(\frac{d}{2\pi}\bigg)^{\frac{d}{2}}
\frac{\Gamma(\frac{4-d}{2})}{\Gamma(\frac{8-d}{2})} \,,
\qquad
&
C_2=\frac{d-2}{d}
 \bigg(\frac{d}{2}\bigg)^{\frac{d-2}{2}}
 \frac{2 \Gamma(\frac{6-d}{2})}{\Gamma(\frac{d}{2})\Gamma(\frac{10-d}{2})}
\,,
\end{align}
are positive constants.
They are divergent at some even $d$,
but this comes from the point $\sigma=\sigma'$ in the integral
and the divergence is an artifact due to using a continuum variable.
In evaluating the size, what we need is not the precise values of $C_1$ and $C_2$,
but the fact that they are $N$-independent positive constants.
We shall regard them simply as positive constants and omit them from the expression in any case.

The consistency condition for UEM is that the size should be given by the free walk
of the fundamental bond size $\ell'$, $\vev{\mathbf{R}^2}=N \ell^{\prime 2}=N a^2 \ell^2$;
namely the parameter $a$ is chosen so that the second, third and forth terms
in \eqref{eq:108} cancel out.
The consistency condition becomes
\begin{align}
  \label{eq:118}
  a^d-a^{d+2} +  u N^{\frac{4-d}{2}}
- g^2 a^{2} N^{\frac{6-d}{2}} =0 \,,
\end{align}
where we have dropped unimportant numerical coefficients.
Note that if $g^2=u=0$, we have essentially a unique solution $a=1$ and
we recover the free walk result.
In the following, we shall find the solution of $a$ in various situations.
First, a pure repulsive case ($u>0, g=0$) is discussed to be capable of realizing
expanded configurations larger than $R_0$;
especially Flory's exponent for real polymers is successfully reproduced.
We briefly mention that UEM is not suitable for a pure gravitational case ($g>0, u=0$),
and then move on to the discussion of the generic case ($g,u>0$)
where a size scaling similar to the one in the variational method is found.

We start with the case with no gravitational interaction,
namely a real polymer ($g^2=0$, $u > 0$).
The consistency condition is 
 $ a^2 = 1 + u a^{-d} N^{\frac{4-d}{2}}$.
If the repulsive force is very weak  ($u N^{\frac{4-d}{2}} \ll 1$), 
we may take the leading order in $u$, and find
$R^2 \simeq  N\ell^2\big(1  + u N^{\frac{4-d}{2}} \big)$.
Since $N \gg 1$, as we have discussed at the beginning of this section,
the size is sensitive to the value of $u$ only for $d<4$.
On the other hand, for $u N^{\frac{4-d}{2}} \gg 1$, the condition
leads to
\bea
a^2 \simeq u^{\frac{2}{d+2}} N^{\frac{4-d}{d+2}} \,,
\label{eq:a_UEM_repulsive}
\eea
and thus we find
\bea
R^2  \simeq \ell^2 u^{\frac{2}{d+2}}  N^{\frac{6}{d+2}}
\label{eq:size_UEM_repulsive}
\,.
\eea
At the marginal coupling $u_o\sim N^{\frac{d-4}{2}}$,
the size becomes $R_0$.
When $u \sim O(1)$, we have
the relation $R \simeq \ell  N^{\frac{3}{d+2}}$ which is known as Flory's exponent
for real polymers \cite{Flory}.
As just mentioned, this result is valid for $d\leq 4$,
and we have witnessed that UEM is capable of evaluating a scaling size
which is larger than that of free walks.

Next, we consider the case with pure attractive force ($u=0$, $g^2 > 0$).
The self-consistency relation is
  $a^2 = 1-g^2a^{2-d} N^{\frac{6-d}{2}}$.
For $g^2 N^{\frac{6-d}{2}} \ll 1$, we observe a perturbative
correction to the size, and find
$R^2 \simeq N\ell^2 \big(1- { g^2} N^{\frac{6-d}{2}}\big)$.
From this relation, the marginal coupling $g_o$ again appears to be
$g_o \sim N^{\frac{d-6}{4}}$ \cite{Horowitz:1997jc}.
Since $a$ has to be positive, there does not exist a solution
at strong coupling $g^2 N^{\frac{6-d}{2}} \gg  1$ (namely $g > g_o$).
Thus, UEM fails to reproduce the scaling by Horowitz and Polchinski
\cite{Horowitz:1997jc}.
If we switch to a repulsive long range force, $g^2 <0$, we find
\begin{align}
a^2 \simeq (-g^2)^{\frac{2}{d}} N^{\frac{6-d}{d}} \,,
\qquad
R^2  \simeq 
(-g^2)^{\frac{2}{d}} \ell^2 N^{\frac{6}{d}} \,.
\end{align}
For $|g^2| \sim O(1)$,
this result agrees with Flory type of scaling argument for
a polyelectrolyte (a single charged polymer) \cite{Pfeuty}, though it  
is not exactly the same as that of renormalization group analysis
$R \simeq \ell N^{\frac{2}{d-2}}$ \cite{DeGennes, Pfeuty}.

Finally, we consider a general case ($g^2>0,u >0 $).
The number of terms in the consistency condition \eqref{eq:118} that come into balance at the leading order
in $N$ would vary.
As in the analysis of the variational method, we take $u$ as a given value and observe
the size change as a function of $g$.
Let us first consider the situation in which the repulsive force is effective, 
$u>u_o$.
When $g$ is small enough, $a$ and the size $R$ are given by \eqref{eq:a_UEM_repulsive}
and \eqref{eq:size_UEM_repulsive} respectively.
The size starts to change when $g$ term in \eqref{eq:118} is comparable to the second and the third term as
\bea
a^{d+2} \sim uN^{\frac{4-d}{2}} \sim g^2 a^2 N^{\frac{6-d}{2}} .
\label{balance_UEM_1}
\eea 
This leads to a marginal coupling $g_o'' \simeq u^{\frac{d}{2(d+2)}} N^{-\frac{3}{d+2}}$ 
at which $a$ starts to decrease; for larger values of $g$,
it is easy to see that
the last two terms in \eqref{eq:118} (or \eqref{balance_UEM_1}) are dominant.
This determines $a$ as
\begin{align}
  a \simeq \sqrt{\frac{u}{g^2 N}} \,,
\label{eq:a_sol_gen}
\end{align}
which leads to the mean size squared
\begin{align}
  R^2 = \ell^2 a^2 {N} \simeq  \frac{\ell^2 u}{g^2} \,.
\label{eq:ms_size_UEM_gen}
\end{align}
Note that this takes the same form as in the harmonic potential analysis
\eqref{eq:ms_size_harmonic_gen}.
Thus, when $u> u_o$, the size begins with an expanded size
$R \simeq \ell u^{\frac{1}{d+2}}  N^{\frac{3}{d+2}}$ for small $g$,
and as increasing $g$ the configuration starts to shrink as $R \simeq \ell \sqrt{u}/g$ at $g=g_o''$, and
eventually collapses to a black hole of size $R_c$ at $g=g_c'$ where the critical size $R_c$ and the coupling $g_c'$ are defined in \eqref{eq:critical_radius} and \eqref{eq:critical_coupling} respectively.
(This analysis provides $u>u_o$ part of Fig.~\ref{fig:phase_2_d_4}
and Fig.~\ref{fig:phase_d_4and5} given in the next section).
Note that if $u > N^{d-1}$, the critical coupling $g_c'$ is larger than the marginal coupling $g_o''$ which implies that the configuration is covered by the horizon even before its size starts
to shrink.

Next, we consider the opposite situation, $u < u_o$.
For a small value of $g$, $a^d$ and $a^{d+2}$ terms in \eqref{eq:118}
give a dominant balance solution $a=1$.
Thus the size starts with $R_0$.
As $g$ increases, the term $g^2 a^2 N^{\frac{6-d}{2}}$ eventually becomes $O(1)$
at $g = g_o$, and the solution of $a$ starts to change.
Since $uN^{\frac{4-d}{2}} \ll O(1)$, the repulsive force does not participate
in the balance condition unless $N$ dependence of $a$ changes as $a \ll 1$.
However, before the repulsive force becomes effective,
the size behavior is the same as the pure attractive force case ($g^2>0$,
$u=0$) 
$R^2 \simeq N\ell^2 \big(1- { g^2} N^{\frac{6-d}{2}}\big)$ 
which has already been analyzed, and UEM fails to provide a consistent
size decreasing behavior as discussed there.
Therefore, we conclude that the UEM analysis is not reliable when
$u< u_o$, and we use the result of the variational method in this region instead.

\subsection{Phase diagrams}
\label{sec:phase-diagram}

In this subsection, we organize the size behavior analyzed so far into ``phase diagrams''
in various dimensions.
As discussed in the introduction, the shape of polymers is determined by the confrontation
among four types of self-interactions; the entropic diffusive and elastic forces, 
the repulsive interaction, and Newton gravity.
Two of them, the diffusive and the repulsive forces, make the size of configuration
larger, while the other two, the elastic and Newton forces, are to compress the configuration.
Thus, typically, there are four types of configuration that are characterized by the balance
between two out of these four interactions; one from the former expanding forces
and another from the latter attracting forces.
These regions have different size scalings that depend on $g$, $u$, and $N$,
and are separated by
boundary lines that are determined by the conditions which two forces
coming to balance.
We call this diagram simply a ``phase diagram.''
The boundary lines are parametrized by various marginal couplings,\footnote{%
Another marginal coupling $g_o' \simeq \sqrt{\frac{u}{N}}$, at which
gravity starts to change the size against the repulsive force,
is also important but does not appear as a boundary line.}
$u_o \sim N^{\frac{d-4}{2}}$ (the repulsive force becomes effective against entropic elasticity),
$g_o \sim N^{\frac{d-6}{4}}$ (gravity starts to balance with entropic diffusive force),
$\tilde{g}_o\simeq u^{\frac{d-4}{2(d-2)}} N^{-\frac{1}{d-2}}$
(gravity, the repulsive, and entropic forces are in equilibrium),
and $g_o''$ (gravity, the repulsive, and entropic forces come to balance in an expanded configuration).
There also exists a special region in which the whole configuration goes inside
the Schwarzschild radius of a black hole of the same mass; we call it a ``black hole'' phase.
The boundary of this domain is given by either of the critical couplings, 
$g_c \sim N^{-\frac{1}{2}}$ (collapse against the entropic force)
or $g_c' \simeq u^{\frac{d-2}{2d}} N^{-\frac{1}{d}}$ (collapse against the repulsive force).

\begin{figure}[tbh]
  \centering
  \includegraphics[width=12cm]{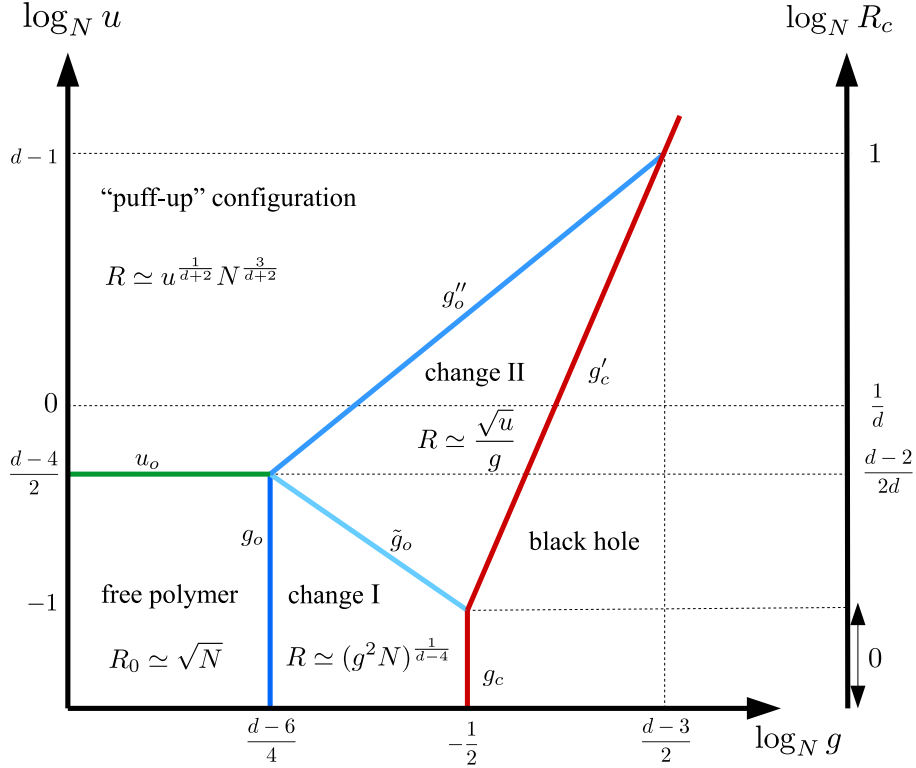}
  \caption{The phase diagram in $2<d<4$. We set $\ell=1$.
The right vertical axis shows the size of a corresponding black hole
for a given value of $u$.}
  \label{fig:phase_2_d_4}
\end{figure}

Fig.~\ref{fig:phase_2_d_4} shows the phase diagram for $2< d< 4$.
The horizontal and the vertical axes are $\log_N g$ and $\log_N u$ respectively,
and $N$ independent factors are neglected.
There are five domains that we have just discussed, and
we look at each of them in more detail in order.
On the left--bottom corner, for small $u$ and $g$, there is a ``free polymer''phase
whose size is given by $R_0 \simeq \ell \sqrt{N}$ in any dimension.
In this domain, the diffusive and the elastic entropic forces balance out, and the size
is stable against the change of $u$ and $g$ until they reach certain marginal values.
Going up vertically,
the configuration starts to expand at $u_o$,
and we come into the region that we call ``puff-up configurations,''
where the repulsive force and the entropic elasticity balance out.
The boundary is given by $u_o$.
The size is given by $R \simeq \ell (uN^3)^{\frac{1}{d+2}}$ which is
stable against the change of $g$ but changes as $u$ varies.
We return to the free polymer domain and go to right as $g$ gets larger
(in the $u < u_o$ region).
It is not difficult to check that $g_o < \tilde{g}_o$ for $u < u_o$ in $2< d< 4$,
and Newton force comes to balance with the entropic diffusive force at $g_o$.
On the right of this boundary, we are in a phase called ``change I'' in which the size decreases as $g$ increases, as $R \simeq \ell (g^2 N)^{\frac{1}{d-4}}$,
but is not sensitive to the change of $u$.
This is equivalent to the behavior of free self-gravitating polymers (no repulsive force),
and it may collapse to a black hole at the critical coupling $g_c$.
It indeed happens if $u < N^{-1}$.
When $u > N^{-1}$, $\tilde{g}_o$ is smaller than $g_c$, and Newton force 
and the repulsive force
will balance out before it becomes a black hole.
Thus, after crossing a border line parametrized by a marginal coupling
$g_o'$,
the different size scaling $R \simeq \ell \sqrt{u}/g$ applies, and we call this region
``change II.''
Note that $g_o' < \tilde{g}_o$ and the size immediately starts to change
after crossing the boundary line.
One can also come into change II from a puff-up configuration region, for a given $u > u_o$,
by increasing $g$ larger than $g_o''$.
Thus, the boundary between change I and change II is given by $\tilde{g}_o$,
while the one between Puff-up and change II is by $g_o''$.
If we further increase $g$ in change II region, the configuration will be smaller than
the size of the horizon at the certain critical coupling $g_c'$.

These phase boundaries are shown in Fig.~\ref{fig:phase_2_d_4}.
The green horizontal segment is given by $u_o$ which separates a free polymer 
and a puff-up regions.
Three blue lines (blue, light blue and cyan) 
denotes the point at which Newton gravity participates in force balance.
They are parametrized by $g_o$, $\tilde{g}_o$, and $g_o''$.
The red lines consist of $g_c$ and $g_c'$ at which the size of the horizon
catches up the size of a polymer, and the configuration may collapse into
a black hole.
It is easy to see that $g_c'$ is more steep than $g_o''$ in general, 
and at $\log_N g=\frac{d-3}{2}$ and $\log_N u=d-1$, $g_c'$ becomes smaller than $g_o''$. 
After this point, the configuration may become a black hole even no interaction work to shrink its size.
On the right of the figure, the size of a corresponding black hole $R_c$ for a given $u$
is also shown.
For $u< N^{-1}$, the corresponding point is given by $g_c \sim N^{-\frac{1}{2}}$
and $R_c \simeq \ell$.
If the repulsive force is significant in balance conditions, the size of corresponding
black holes gets enhanced, as $\log_N (R_c/\ell) \sim \frac{1+\log_N u}{d}$
(see \eqref{eq:critical_radius}), and black holes swell in general.

\begin{figure}[htb]
  \centering
  \includegraphics[width=12cm]{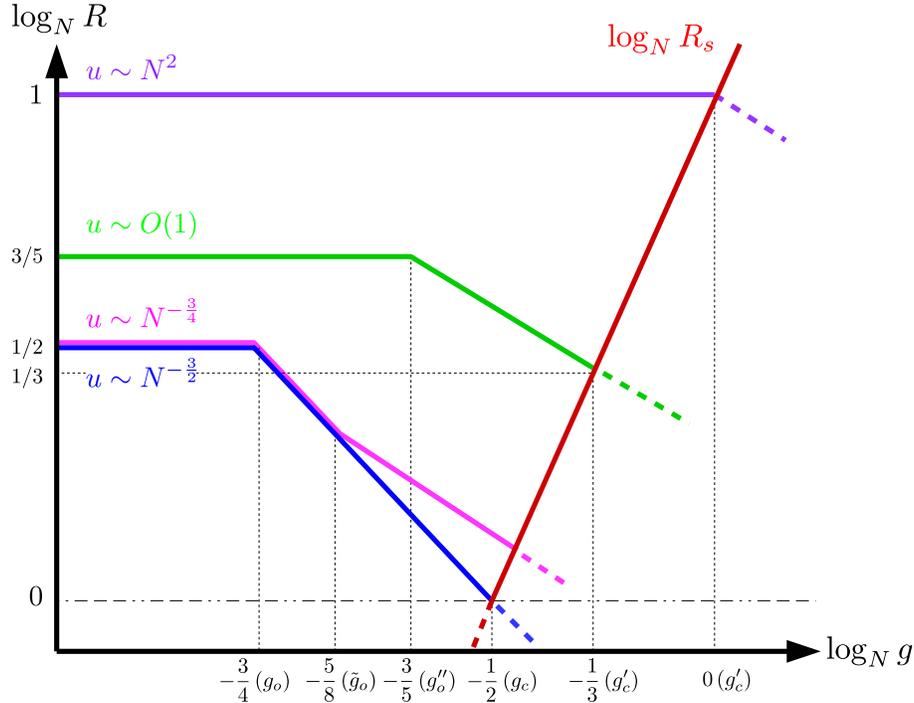}
  \caption{The change of the size for various values of $u$ in $d=3$.
$\ell$ is set to be 1.
 The red line
represents the horizon size at $g$.
On the horizontal axis, we show which marginal or critical couplings are relevant
for each numbers. Note that $\tilde{g}_o$, $g_c''$, and $g_c'$ are $u$ dependent.}
  \label{fig:size_change_d=3}
\end{figure}
Fig.~\ref{fig:size_change_d=3} shows the size change with respect to
$\log_N g$ for various values of $u$ in $d=3$.
This figure is schematic and the scalings are adjusted to draw the diagram.
We pick up four typical values of $u$; $u \sim N^{-\frac{3}{2}}$ (the repulsive force
does not work throughout), $u \sim N^{-\frac{3}{4}}$ (the configuration experiences both
change I and II), $u \sim N^0$ (a real polymer whose largest size is given by Flory's scaling),
and $u \sim N^2$ (the horizon size catches up before the interaction starts to work).
The red line here shows the size of a horizon for the same mass.
The crossing points with each line determine the corresponding points.

\begin{figure}[htb]
  \centering
  \includegraphics[width=7.5cm]{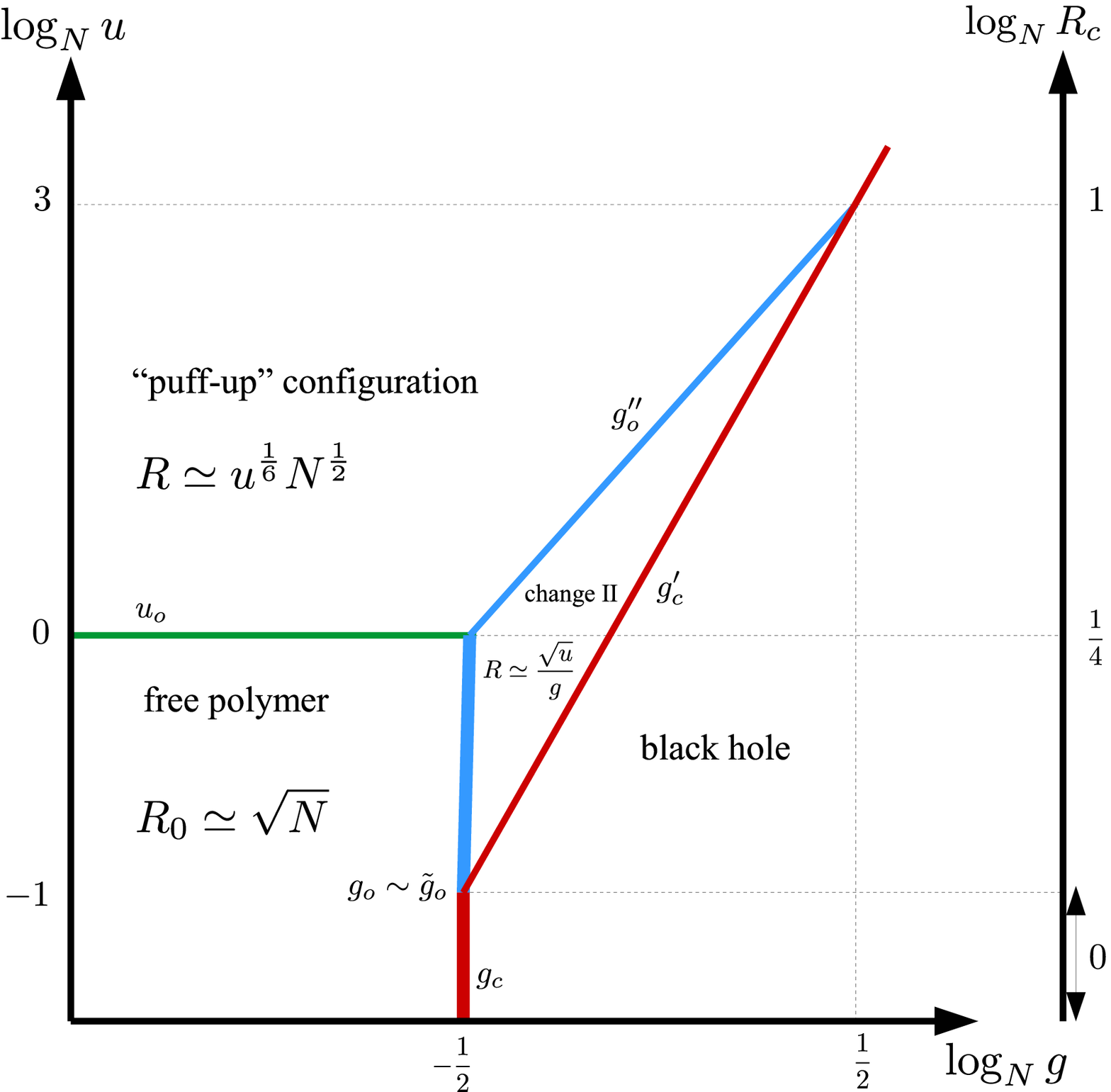}
\hspace{1em}
  \includegraphics[width=8cm]{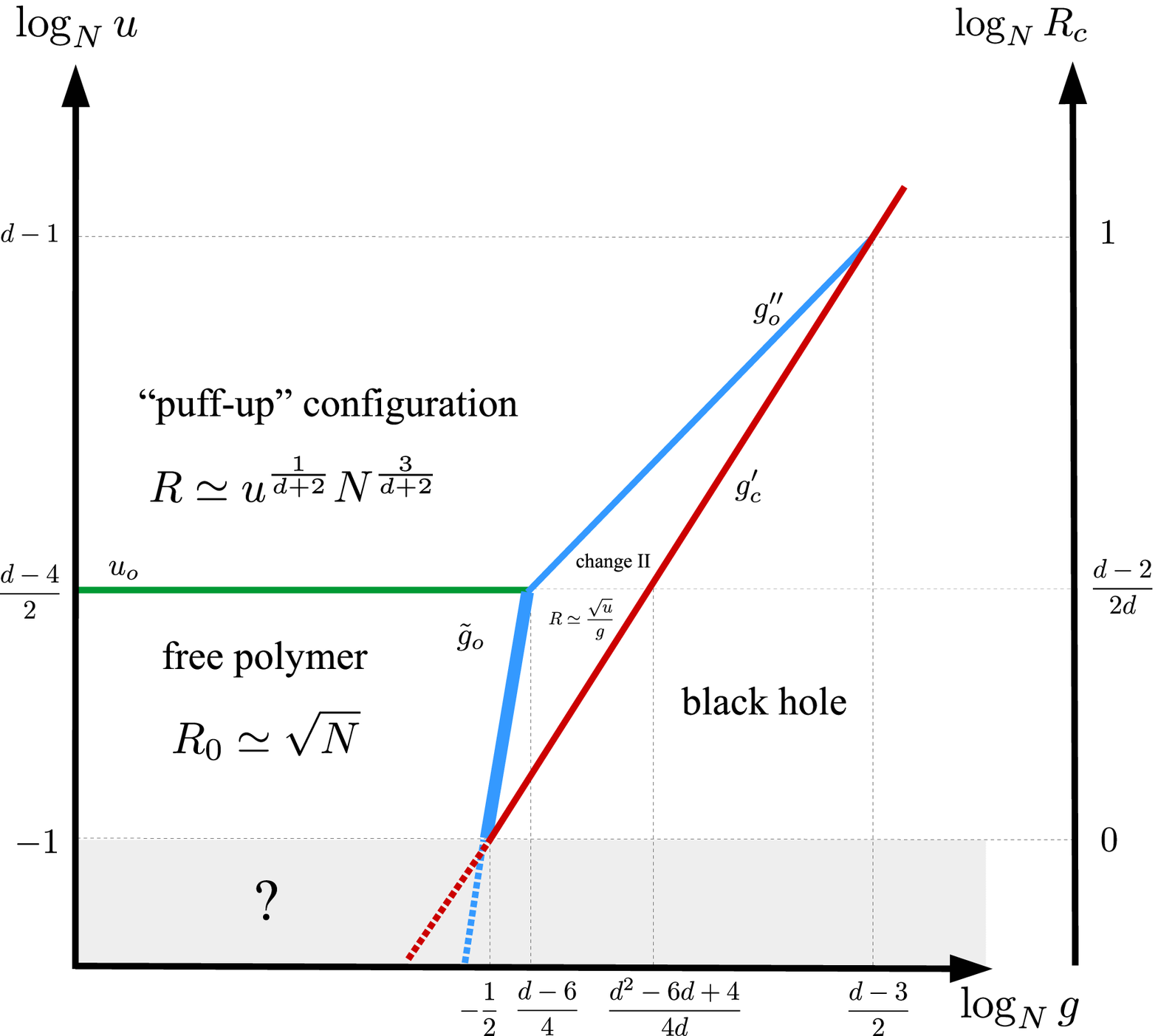}
  \caption{(Left) The phase diagram in $d=4$. The vertical thick lines (with two colors)
denotes the point where the size of configuration jumps.
(Right) The phase diagram in $d>4$. Below $u \sim N^{-1}$, the behavior of the size is rather obscure. In both diagrams, $\ell$ is set to be 1.}
  \label{fig:phase_d_4and5}
\end{figure}
Next, we quickly examine the cases in $d=4$ and $d>4$.
From Fig.~\ref{fig:phase_2_d_4}, one can see the point at which four different regions meet
(at $(\log_N g, \log_N u)=(\frac{d-6}{4}, \frac{d-4}{2})$) moves to a right-up direction
as $d$ comes closer to 4.
At $d=4$, some boundary lines merge and we obtain the left diagram in Fig.~\ref{fig:phase_d_4and5}.
The change I region disappears, and there are four regions left.
For $u< u_o \sim N^0$, the size is given by $R_0$ in a weak coupling region.
As $g$ gets larger, it may collapse into a black hole if $u < N^{-1}$
or may start to change its size as $R \simeq \ell \sqrt{u}/g$.
A crucial difference from the situation in $2<d<4$ is that for $u < u_o$, by crossing
a vertical boundary line (denoted by thick light blue and red lines),
the size of a configuration jumps; it suddenly collapse to a black hole of size $\ell$ 
if $u< N^{-1}$, or it quickly shrinks down to the size $R$ determined by the value of $u$
for $N^{-1}<u<N^0$.
For $u> u_o$, the repulsive force is significant, and the behavior is analogous to
the case in $2<d<4$, and the size smoothly starts to change by crossing the
boundary line $g_o''$.
In $d>4$, the diagram is given by the right in Fig.~\ref{fig:phase_d_4and5}.
It is similar to the situation in $d=4$, but in this case the size scaling
for $u < N^{-1}$ is ambiguous. As discussed in Sec.~\ref{sec:harmonic},
the size scaling shows an anomalous behavior if the repulsive force is turned off.
In $u< N^{-1}$, the repulsive force is not effective and the situation is similar to
that pure gravitating case. 
The other part is close to the $d=4$ diagram.

\section{Conclusion}
\label{sec:conlusion-discussion}


In this paper, we have evaluated the averaged size of a long polymer which interacts
with itself through Newton gravity and also has self-avoiding
property that is implemented by a contact repulsive interaction,
in spatial $d$ dimensions.
The mathematical description for
the statistical property of polymers is given by a self-avoiding random walk with self-interactions,
and is analyzed by an effective Hamiltonian \'a la Edwards.
We have evaluated the expectation value of the end-to-end distance squared with respect to this Hamiltonian by employing two approximations; a variational analysis by a trial Hamiltonian
with a harmonic potential and a uniform expansion model.
The size of the polymer changes as the strength of the interactions varies.
It has been found that the variational calculation gives a reasonable result when the size gets smaller than that of the free random walk $R_0 \simeq \ell N^{1\over2}$, 
in which the attractive force overcomes the repulsive 
or entropic diffusive force.
This method, however, fails to explain the case where the configuration expands; even if the repulsive force becomes dominant, the size remains as $R_0$.
This may be a limitation of the approximation due to using a harmonic potential to describe a dominant configuration.
On the other hand, the uniform expansion model, where the configuration is fixed to remain the free walk 
and the change of the size is encoded in a renormalization of the
length of the bonds between monomers, turns out to be capable of explaining reasonable
changes of the size in the expansion case, where
the repulsive force is sufficiently strong.
As the size decreases, this method does not work at strong coupling for the pure attractive case,
but it provides a consistent behavior when Newton and the repulsive forces balance out.
In the most compelling case where the two forces are in balance, 
the size is found to be
\begin{align}
  R \simeq \ell \sqrt{\frac{u}{g^2}} \,,
\end{align}
where $g$ and $u$ stand for the coupling constants of Newton and the repulsive forces respectively,
and this scaling is valid for both contraction and expansion cases.
The size scaling varies as which two of four competing self-interactions are in balance,
and there in general exist four scaling regions depending on the values of $g$ and $u$.
Together with a situation in which the whole configuration is covered by the Schwarzschild radius
of a corresponding black hole, unified picture of the size behavior is shown in
the phase diagrams Fig.~\ref{fig:phase_2_d_4} and \ref{fig:phase_d_4and5}.

This analysis is motivated by a conjecture by Susskind \cite{Susskind:1993ws}
on a correspondence
between long fundamental strings and a small black hole,
and
it is thus of interest to understand the behavior of the size of a long string,
especially when it contracts.
In the case of pure Newton gravity,
Khuri\cite{Khuri:1999ez} has carried out a variational analysis for Edwards Hamiltonian,
and derived a scaling consistent with the result of Horowitz and Polchinski \cite{Horowitz:1997jc}.
In this paper, a generalization with the repulsive force is considered, 
and we have estimated the critical coupling and the critical size, at which a string may be transformed into a corresponding black hole.
The repulsive force pushes the critical gravitational coupling to
larger values, and the size of a black hole at the transition point will swell,
if the repulsive force is sufficiently strong.
Although the origin of the repulsive force in string theory is not transparent, it has been argued that it would emerge nonperturbatively to explain the exponential spreading of string degrees of freedom in the context of black hole complementarity \cite{Susskind:1994vu, Ropotenko:2008ti}, and our analysis may serve an interesting observation when the repulsive force is at work.
There might be other sources to induce effective self-avoiding property;
for example, $B$-field (or other higher rank tensors) exchange or fermionic degenerate force.
The former would not be important for spherically symmetric configurations as
it cancels out under the average, but may be effective if anisotropy is introduced.
The latter plays a central role in the gravitational collapse of stars, and if a polymer carries fermionic degrees of freedom, such as a model of superstring,
the degenerate force can also be important to determine the properties of the polymer when the size is sufficiently small.


We have been concentrating on the size of a configuration throughout this paper.
There are more quantities of interest such as density, or elasticity (or pressure) distribution.
These are information necessary to write down an equation of state.
Once we know these quantities, we can analyze the change of the size with more sophisticated methods which are used for gravitational collapse of stellar objects.
Since a strong gravitational force is required to crush an object if a repulsive force is introduced, 
there might be a region where we need to invoke general relativity to describe the whole process, instead of Newton gravity.
If it is the case, it will be interesting to investigate Tolman-Oppenheimer-Volkoff equation \cite{TOV} for collapsed polymers 
where one can treat the dynamics of the system including background spacetime in a self-consistent way.


Though our study is indeed motivated by the string/black hole correspondence, the analysis will also be interesting from the point of view of polymer physics.
As argued in this paper, it is intriguing that there exist nontrivial scalings for real polymer with a long-range attractive force.
By scaling the attractive coupling with respect to $N$, we can determine the scaling exponent for a fixed magnitude of the repulsive force ($u \sim O(1)$ for real polymers).
For example, we take $g \sim N^\alpha$ and $u \sim O(1)$, then the size scales as $R \sim N^{-\alpha}$ when Newton and the repulsive forces are in balance.
It has been known that these exponents will
get modified if we employ renormalization group analysis \cite{DeGennes}.
It will also be interesting to carry out renormalization group analysis in the current model
to obtain more accurate exponents.

A related observation concerns the size at the corresponding point for real polymers with $u \sim O(1)$.
The size coincides with that of close-packed configurations in $d$ dimensions, $N^{1\over d}$, which is the smallest possible size scaling for $N$-step SAW.
In the case of real polymers, the fundamental molecule structure may be
broken down and the mathematical description may not be valid.
However, it is curious that this scaling may be just a coincidence, or 
a more profound reason exists behind it.
If Newton gravity is not valid before the configuration reaches this size, it is interesting to check whether this smallest size scaling still appears or not in analysis with general relativity.

Another interesting issue may be about the symmetry of configurations.
Although the statistical average leads to spherically symmetric configurations, 
each snapshot of configurations is known to be aspherical \cite{asphericity}.
Once interaction becomes effective, the shape will be further distorted.
Thus, the gravitational collapse of polymers should have much richer contents than
those discussed here, which may deserve further investigation.

We hope to revisit these issues in future and to report elsewhere.

\section*{Acknowledgment}
\label{sec:acknowledgement}

The authors thank K.~Hashimoto, T.~Houri, 
N.~Ishibashi, T.~Kuroki, M.~Shigemori, K.~Yamada and Y.~Yasui
 for useful discussions and valuable comments.
They also thank Kin-ya Oda for useful suggestions.
The work of S.~K. is supported by 
NSC103--2811--M--033--004
and NSC 103-2119-M-007-003.
S.~K. would like to thank NCTS, Physics Division where a part of the work
has been carried out.
The work of T. M is supported in part by JSPS KAKENHI Grant Number 26400258.

\appendix
\section{Miscellaneous Calculations}
\label{sec:misc-calc}

In this appendix, we present some details of the evaluation of each term
in \eqref{eq:UEM_VEV1} and \eqref{eq:UEM_VEV2}.
We start with the ones involving the kinetic term,
which contains a derivative.
It is better to come back to a discrete description,
$\int_0^N d\sigma \rightarrow \sum_{n=1}^N$ and
$\frac{\partial \mathbf{R}(\sigma)}{\partial
  \sigma}\rightarrow \mathbf{R}_n-\mathbf{R}_{n-1}$
(for $\sigma \in (n-1,n]$), where
we use a subscript to denote the position of $n$-th monomer.
Thus,
\begin{align}
\frac{1}{\ell^2} \int_0^N d\sigma \, 
\vev{ \left( \frac{\partial \mathbf{R}}{\partial \sigma}\right)^2}'
=&
\frac{1}{\ell^2} \sum_{n=1}^N \, \vev{(\mathbf{R}_n - \mathbf{R}_{n-1})^2}'
\nn\\=&
\frac{1}{\ell^2} \sum_{n=1}^N  
\int \mathcal{D}\mathbf{R}_{n-1} \mathcal{D}\mathbf{R}_{n} 
\mathcal{D}\mathbf{R}_{N}
G'(0,n-1)G'(n-1,n),G'(n,N) 
(\mathbf{R}_n - \mathbf{R}_{n-1})^2
\nn\\=& a^2 N
\,,
\end{align}
and
\begin{align}
&
\frac{1}{\ell^2} \int_0^N d\sigma \, 
\vev{ \left( \frac{\partial \mathbf{R}}{\partial \sigma}\right)^2
(\mathbf{R}(N)-\mathbf{R}(0))^2 }'
\nn\\=&
\frac{1}{\ell^2} \sum_{n=1}^N 
\bigg\langle
(\mathbf{R}_n - \mathbf{R}_{n-1})^2
\big[
(\mathbf{R}_N - \mathbf{R}_{n})
+(\mathbf{R}_n - \mathbf{R}_{n-1})
+\mathbf{R}_{n-1} \big]^2  \bigg\rangle'
\nn\\=&
\frac{1}{\ell^2} \sum_{n=1}^N   
\bigg[
\vev{(\mathbf{R}_N - \mathbf{R}_{n})^2}'
\vev{(\mathbf{R}_n - \mathbf{R}_{n-1})^2}'
+\vev{(\mathbf{R}_n - \mathbf{R}_{n-1})^4}'
+\vev{(\mathbf{R}_{n-1})^2}'
\vev{(\mathbf{R}_n - \mathbf{R}_{n-1})^2}'
\bigg]
\nn\\=&
a^4 \ell^2 N\left( N+\frac{2}{d} \right) 
\,,
\end{align}
where we have used $\mathbf{R}(0)=\mathbf{0}$ and the properties of Gaussian average.

Next, we move on to the interaction terms.
As before, $V(\sigma,\sigma')$ is symmetric, and we take
$\sigma' > \sigma$
by rewriting the integral as $2 \int_0^N d\sigma' \int_0^{\sigma'} d\sigma$.
It should be noted that we take $\sigma' > \sigma$, not $\sigma' \geq \sigma$.
Since we do not consider the self-interaction of each monomer, the point
$\sigma'=\sigma$ should be excluded.
For a discrete description, we then consider $n' > n$.
The expectation values of the delta function part are
\begin{align}
  \label{eq:101}
 \vev{\delta^{(d)}(\mathbf{R}(\sigma)-\mathbf{R}(\sigma'))}'
=&
\int \mathcal{D}\mathbf{R}(\sigma) \mathcal{D}\mathbf{R}(\sigma') 
\mathcal{D}\mathbf{R}(N)
G'(0,\sigma)G'(\sigma,\sigma')G'(\sigma',N) 
\delta^{(d)}(\mathbf{R}(\sigma)-\mathbf{R}(\sigma'))
\nn\\=&
\left(
\frac{d}{2\pi a^2\ell^2 (\sigma'-\sigma)}
\right)^{\frac{d}{2}}
\,,
\end{align}
and
\begin{align}
\vev{
(\mathbf{R}_N-\mathbf{R}_0)^2
\delta^{(d)}(\mathbf{R}(\sigma)-\mathbf{R}(\sigma'))}'
=&
a^2 \ell^2 \big(N-\sigma'+\sigma \big)
\left(
\frac{d}{2\pi a^2 \ell^2 (\sigma'-\sigma)}
\right)^{\frac{d}{2}}
\,.
\end{align}

The Newton interaction part is again evaluated by converting to the exponential form \eqref{eq:Coulomb_exp},
\begin{align}
  \label{eq:103}
\vev{  \frac{-g^2 \ell^{d-2}}{|\mathbf{R}(\sigma)- \mathbf{R}(\sigma')|^{d-2}}}'
=&
-g^2 \ell^{d-2}
\frac{i^{\frac{d-2}{2}}}{\Gamma(\frac{d-2}{2})}
\int_0^\infty x^{\frac{d}{2}-2} dx \,
\vev{e^{-ix (\mathbf{R}(\sigma) - \mathbf{R}(\sigma'))^2} }'
\nn\\=&
-g^2 \ell^{d-2}
\frac{i^{\frac{d-2}{2}}}{\Gamma(\frac{d-2}{2})}
\int_0^\infty x^{\frac{d}{2}-2} dx \,
\bigg(1+\frac{2i a^2 \ell^2 (\sigma'-\sigma)}{d} x \bigg)^{-\frac{d}{2}}
\nn\\=&
\frac{-g^2}{\Gamma(\frac{d}{2})}
\bigg(\frac{d}{2a^2}\bigg)^{\frac{d-2}{2}} (\sigma'-\sigma)^{\frac{2-d}{2}}
 \,.
\end{align}
For the one with $(\mathbf{R}(N)-\mathbf{R}(0))^2$ insertion,
the expectation value is
\begin{align}
  \label{eq:106}
&  \vev{e^{-ix (\mathbf{R}(\sigma) - \mathbf{R}(\sigma'))^2}(\mathbf{R}(N)-\mathbf{R}(0))^2}'
\nn\\=&
a^2 \ell^2
\bigg[
N-\sigma'+\sigma
+(\sigma'-\sigma)\bigg(1+\frac{2ia^2 \ell^2
  (\sigma'-\sigma)}{d}x \bigg)^{-1}
\bigg]
\bigg(1+\frac{2ia^2\ell^2 (\sigma'-\sigma)}{d}x \bigg)^{-\frac{d}{2}}
\,.
\end{align}
Upon $x$ integral, we find
\begin{align}
  \label{eq:107}
\vev{\frac{-g^2 \ell^{d-2}}{|\mathbf{R}(\sigma)- \mathbf{R}(\sigma')|^{d-2}} 
(\mathbf{R}(N)-\mathbf{R}(0))^2}'
=&
\frac{-g^2 a^2 \ell^2}{\Gamma(\frac{d}{2})}
\bigg(\frac{d}{2a^2}\bigg)^{\frac{d-2}{2}} (\sigma'-\sigma)^{\frac{2-d}{2}}
\bigg[
N + (\sigma'-\sigma)\frac{2-d}{d}
\bigg]
 \,.  
\end{align}
By collecting these results, we obtain \eqref{eq:R2_UEM}.

\section{A generic power potential and van der Waals interaction}
\label{sec:van-der-waals}

The methods used in this paper can be applied to any power-law long-range force.
We first present a general formula and argue a potential problem that arises
for interactions with higher inverse-power.
As an example, we see that the size scaling due to van der Waals inverse-sextic potential of a real polymer in three dimensions may have difficulty.

We consider a generic power-law potential term in $d$ dimensions,
\begin{align}
  V_\alpha =& 
\frac{-\xi \ell^{\alpha}}{|\mathbf{R}(\sigma)- \mathbf{R}(\sigma')|^{\alpha}}
\,,
\end{align}
where $\xi$ is a dimensionless coupling (positive for an attractive case).
$\alpha=d-2$ corresponds to Coulomb-type interactions, including Newton gravity.
For variational calculation, we evaluate the expectation value of this interaction term
by use of a harmonic trial Hamiltonian,
\begin{align}
  \vev{V_\alpha}_0 =&
-\xi \frac{\Gamma \big(\frac{d-\alpha}{2} \big)}{\Gamma \big(\frac{d}{2} \big)}
\bigg(\frac{qd}{2 F_1(\sigma, \sigma';q)}   \bigg)^{\frac{\alpha}{2}}
\,,
\end{align}
where $F_1(\sigma, \sigma';q)$ is given in \eqref{eq:F1}.
On the other hand, in the uniform expansion model, we need to evaluate the following two terms with Gaussian Hamiltonian \eqref{eq:Edwards_UEM},
\begin{align}
  \vev{V_\alpha}' =& 
-\xi \frac{\Gamma \big(\frac{d-\alpha}{2} \big)}{\Gamma(\frac{d}{2})}
\bigg(\frac{d}{2a^2}\bigg)^{\frac{d-2}{2}} (\sigma'-\sigma)^{-\frac{\alpha}{2}}
\,,\\
\vev{V_\alpha 
(\mathbf{R}(N)-\mathbf{R}(0))^2}'
=&
-\xi a^2 \ell^2 \frac{\Gamma \big(\frac{d-\alpha}{2} \big)}{\Gamma(\frac{d}{2})}
\bigg(\frac{d}{2a^2}\bigg)^{\frac{\alpha}{2}} (\sigma'-\sigma)^{-\frac{\alpha}{2}}
\bigg[
N - (\sigma'-\sigma)\frac{\alpha}{d}
\bigg]
 \,.  
\end{align}
In these calculations, one would have faced divergent integrals for some large values of
$\alpha$.
This is due to a short distance singularities $|\mathbf{R}(\sigma) - \mathbf{R}(\sigma')|\rightarrow0$, but phenomenologically these singularities are avoided
by using an effective repulsive force (thus a more realistic phenomenological potential
is of Lennard-Jones type).
We neglect such divergence and take only $q$ and $a$ dependence.

The variational calculation for the optimized value of $q$ leads to
\begin{align}
  1- N \xi q^{\frac{\alpha-2}{2}} + N u q^{\frac{d-2}{2}}=0 \,,
\end{align}
while the UEM consistency condition is given by
\begin{align}
  1-a^2 + ua^{-d} N^{\frac{4-d}{2}} -\xi a^{-\alpha} N^{\frac{4-\alpha}{2}} =0 \,,
\end{align}
where we have dropped positive numerical coefficients as before.
If the repulsive force is absent (namely $u=0$), the variational calculation
is more reliable, and the optimal value of $q_0$ and the scaling size are given by
$q_0 \simeq (N \xi)^{-\frac{2}{\alpha-2}}$ and 
$R \simeq \ell (N \xi)^{\frac{1}{\alpha-2}}$ respectively.
The size increases as $\xi$ gets larger if $\alpha>2$ and it is an unreasonable behavior.
In the case of Newton gravity ($\alpha=d-2$), this corresponds to $d>4$, as we have observed in
Sec.~\ref{sec:harmonic}.
If the repulsive force is turned on and is balanced with Newton force, 
both methods lead to the same size scaling
\begin{align}
    R \simeq \ell \bigg( \frac{u}{\xi} \bigg)^{\frac{1}{d-\alpha}} \,.
\end{align}
Therefore, if $\alpha>d$, the configuration expands if the attractive force gets stronger.
In the case of Newton gravity, this never happens and then we obtain a reasonable scaling
if the repulsive and Newton force balance out.
However, if the inverse-power of the potential is too large, we again face an unreasonable
behavior, and our approximations both fail.

One of the typical example of this pathological behavior is van der Waals interaction
in three dimensions, where $\alpha=6$ and $d=3$.
At this moment, it is not so clear why we fail to observe that van der Waals force, one
of the realistic interactions of real polymers, makes a configuration smaller;
but one possible explanation may be about the validity of
the perturbative treatment of the interaction, as follows.
By recalling a scaling argument
at the beginning of Section \ref{sec:Self-repell},
we find that the marginal van der Waals coupling for a free configuration
is $\xi_o \sim N$, namely an extremely strong coupling, 
and an $O(1)$ coupling does not affect the configuration in three dimensions.
An $O(1)$ van der Waals coupling  becomes effective only 
when the size of the configuration becomes $R \simeq \ell N^{1/3}$.
This represents close-packed configurations, and then van der Waals interaction may be
important only for phases in dense-packing, like crystallization, for real polymers.
Thus, it may not be justified to perturbatively
treat van der Waals interaction as a long-range interaction
 in the context of determining the shape of long polymers.


 \end{document}